\documentclass[10pt,twocolumn,prl,aps,amssymb,amsmath,tightenlines,showpacs]{revtex4}

\usepackage{graphicx}

\newcommand{\cP}{\ensuremath{\mathcal{P}}}
\newcommand{\cT}{\ensuremath{\mathcal{T}}}

\begin{document}

\title{Response to Shalaby's Comment on ``Families of Particles
with Different Masses in ${\cal PT}$-Symmetric Quantum Field Theory''}

\author{Carl M. Bender$^a$}\email{cmb@wustl.edu}
\author{S. P. Klevansky$^b$}\email{spk@physik.uni-heidelberg.de}
\affiliation{$^a$Physics Department, Washington University, St. Louis, MO 63130,
USA}
\affiliation{$^b$Institut f\"ur Theoretische Physik, Universit\"at Heidelberg,
Philosophenweg 19, 69120 Heidelberg, Germany}

\date{\today}

\begin{abstract}
In a recent Comment [arXiv: 1101.3980] Shalaby criticised our paper ``Families
of Particles with Different Masses in ${\cal PT}$-Symmetric Quantum Field
Theory.'' On examining his arguments, we find that there are serious flaws at
almost every stage of his Comment. In view of space and time considerations, we
point out the major flaws that render his arguments invalid. Essentially Shalaby
is attempting to obtain our results from a variational principle and to find a
physical interpretation of his calculation. The variational procedure that he
uses is inapplicable, and his description of the physics is wrong. We thus
refute his criticism on all levels.

A recent Comment by Shalaby [arXiv: 1101.3980] is wrong.
\end{abstract}


\maketitle

In a recent paper \cite{X1} we conjectured that, while a flavor symmetry group
is conventionally introduced to describe families of particles, such families
might arise naturally from the monodromy structure in the complex-field plane
associated with rotation from one Stokes' wedge to another. Shalaby has
submitted a Comment \cite{X2} arguing that Ref.~\cite{X1} is wrong because,
as he states in the abstract, ``the vacuum with unbroken Z2 symmetry has lower
energy than the vacuum with broken Z2 symmetry'' and thus the theory ``will
prefer to live with the vacuum of unbroken Z2 symmetry in which the proposed
theory is Hermitian.'' The purpose of the current paper is to show that
Shalaby's argument is faulty.

The conjecture in Ref.~\cite{X1} stems from earlier work \cite{X3,X4} in which
it was shown that a quantum-mechanical Hamiltonian can have many different and
independent spectra depending on the choice of complex boundary conditions
satisfied by the eigenfunctions. Surprisingly, if the Hamiltonian is ${\cal PT}$
symmetric then, even though the boundary conditions are imposed in the complex
plane, the eigenvalues can be entirely real and positive. For example, the
Hamiltonian $H=p^2+x^6$ has two real spectra. First, there is the conventional
positive discrete spectrum obtained by requiring the eigenfunctions to vanish as
$|x|\to\infty$ on the real-$x$ axis. Second, there is a different and
unconventional positive discrete spectrum obtained by requiring that the
eigenfunctions vanish as $|x|\to\infty$ inside a pair of Stokes wedges lying
below the positive-real and the negative-real axes. A graph of the
unconventional spectrum is shown in Fig.~1 in Ref.~\cite{X3}. The unconventional
version of the $x^6$ theory is interesting because the theory is not parity
symmetric and thus the one-point Green's function does not vanish.

It was subsequently understood that when the unconventional spectrum is real,
the theory defined by the Hamiltonian and associated boundary conditions is a
{\it Hermitian} theory, but with an unconventional definition of the Hermitian
adjoint \cite{X5,X6}. Since both theories, the conventional one and the
unconventional one, are Hermitian, Shalaby's statement above is misleading.
Furthermore, the claim that the theory prefers to live in one Stokes' wedge
rather than another because the vacuum energy is lower in one wedge than the
other is wrong because these are two independent noninteracting theories. This
would be like saying that the anharmonic oscillator $H=p^2+x^4$ prefers to be
the harmonic oscillator $H=p^2+x^2$ because its vacuum energy is lower.

A complete catalog of all the mistakes in Shalaby's Comment would require many
pages and would be well beyond the scope of this response. Here, we mention just
a few fatal flaws in Shalaby's argument.

Shalaby's Comment rests on a variational calculation that is invalid. A crucial
error in this Comment is the assumption that ``any variational calculation of
the vacuum energy should be higher than the true (exact) vacuum energy'' (ninth
line of Abstract). For ${\cal PT}$-symmetric Hamiltonians this assumption is
false because the Hamiltonian is not conventionally Hermitian. It has been known
for many years that variational approximants for the ground-state energy of a
${\cal PT}$-symmetric Hamiltonian are not necessarily greater than the exact
ground-state energy and that higher variational approximations do not converge
in a monotone fashion to the exact ground-state energy \cite{X4,X7}. This wrong
assumption entirely invalidates Shalaby's argument. In general, to perform such
a variational calculation on a ${\cal PT}$-symmetric Hamiltonian and then claim
that the variational calculation of the ground-state energy lies above the exact
value, one must know in advance the form of the Hermitian adjoint (that is, one
must know the ${\cal C}$ operator).

Apart from this crucial error, there are many further incorrect statements in
the paper. In line four of the Abstract the author claims that the
Dyson-Schwinger equations in quantum field theory ``stem from a variational
principle.'' This is certainly not true; the exact equations of a quantum field
theory stem from an action principle, such as the Schwinger action principle. An
action principle (in which one finds a stationary point in function space) gives
the exact field equation, while a variational principle (for which one minimizes
the energy) relies on a trial wave function in which one evaluates the
expectation value of the Hamiltonian in order to obtain an approximation. The
Dyson-Schwinger equations are an exact system of differential equations that
express the relationships among the Green's functions. These equations cannot be
solved unless they are truncated. Of course, any equation, such as a truncated
Dyson-Schwinger equation can be derived from an ad hoc algorithm that is set up
as a variational calculation. However, Shalaby's ad hoc algorithm, in which he
varies a mass parameter rather than a field, has no physical meaning unless the
variational calculation is performed in a Hermitian setting in which it makes
sense to minimize an energy.

The whole point of Ref.~\cite{X1} is that one needs more than the Hamiltonian
because any given Hamiltonian can have many distinct and unrelated solutions
(and associated energy levels) in different sectors of complex field space, and
these sectors are characterized by the choice of complex boundary conditions in
field space. In the current Comment, the author has no way to impose the
boundary conditions in a coherent manner because the author has not calculated
the expectation value of the Hamiltonian in complex field space. He merely uses
a one-dimensional mass parameter to reproduce the numerical results of Bender
and Klevansky in an ad hoc fashion, and the author's calculational method is
insensitive to the choice of boundary conditions.

Furthermore, the Schwinger-Dyson equations cannot be used to determine the
ground-state energy of a quantum field theory because they are insensitive to
any additive constant in the Hamiltonian. (The Dyson-Schwinger equations can
only be used to calculate energy excitations, and not an absolute energy such as
the ground-state energy.) Thus, the author's entire argument regarding the
ground-state energy is meaningless.

We emphasize that in Ref.~\cite{X1} it is shown that for a $\phi^6$ theory there
are two noninteracting sectors in which the excitations are different. The only
way for a higher-energy state in one sector to decay into a lower-energy state
in another sector is for the sectors to be dynamically coupled (say, by
coupling them to an electromagnetic field). In the model of Ref.~\cite{X1} this
cannot happen because the sectors live in totally different and noninteracting
Stokes wedges. Thus, Shalaby's statement on page 3 that ``the theory does have
one and only one acceptable vacuum solution and thus describe (sic) only one
particle and not a family of particles'' is totally wrong. There are two
noninteracting ground states.

Even if everything that Shalaby had said up to this point were true, his
concluding argument is again wrong. We understand that particles are organized
into different families if we make the approximation that the families are
noninteracting; the electron lives in one sector, the muon in another, and
so on. Each sector in this approximation has its own ground state. Of course,
in nature these sectors are really interacting, and the muon can decay into an
electron. Thus, there really is only one ground state for the entire system.
But, this does not imply that there are not different families of particles, as
Shalaby tries to argue.

It is pointless to press on with our criticism. The substantial flaws in
Shalaby's logic invalidate his paper.

CMB is grateful to the Graduate School at the University of Heidelberg for its
hospitality. CMB thanks the U.S.~Department of Energy for financial support.

\end{document}